# Marangoni droplets of dextran in PEG solution and its motile change due to coil-globule transition of coexisting DNA

Tomohiro Furuki[1,2,†], Hiroki Sakuta[3,4,†], Naoya Yanagisawa[3], Shingo Tabuchi[5], Akari Kamo[5], Daisuke S. Shimamoto[3], Miho Yanagisawa[2,3,4,5*]

[1] Faculty of Pure and Applied Sciences, University of Tsukuba, Tennodai 1-1-1, Tsukuba, Ibaraki 305-8573, Japan

[2] Department of Integrated Sciences, College of Arts and Sciences, The University of Tokyo, Komaba 3-8-1, Meguro, Tokyo 153-8902, Japan

[3] Komaba Institute for Science, Graduate School of Arts and Sciences, The University of Tokyo, Komaba 3-8-1, Meguro, Tokyo 153-8902, Japan

[4] Center for Complex Systems Biology, Universal Biology Institute, The University of Tokyo, Komaba 3-8-1, Meguro, Tokyo 153-8902, Japan

[5] Graduate School of Science, The University of Tokyo, Hongo 7-3-1, Bunkyo, Tokyo 113-0033, Japan

**Abstract**

Motile droplets using Marangoni convection are attracting attention for their potential as cell-mimicking small robots. However, the motion of droplets relative to the internal and external environments that generate Marangoni convection has not been quantitatively described. In this study, we used an aqueous two-phase system (polyethylene glycol (PEG) and dextran) in an elongated chamber to generate motile dextran droplets in a constant PEG concentration gradient. We demonstrated that dextran droplets move by Marangoni convection, resulting from the PEG concentration gradient and the active transport of PEG and dextran into and out of the droplet. Furthermore, by spontaneously incorporating long DNA into the dextran droplets, we achieved cell-like motility changes controlled by coexisting environment-sensing molecules. The DNA changes its position within the droplet and motile speed in response to external conditions. In the presence of $Mg^{2+}$, the coil-globule transition of DNA inside the droplet accelerates the motile speed due to the decrease in the droplet's dynamic viscosity. Globule DNA condenses at the rear part of the droplet along the convection, while coil DNA moves away from the droplet's central axis, separating the dipole convections. These results provide a blueprint for designing autonomous small robots using phase-separated droplets, which change the mobility and molecular distribution within the droplet in reaction to the environment. It will also open unexplored areas of self-assembly mechanisms through phase separation under convections, such as intracellular phase separation.

**Introduction**

Artificial cells have been extensively investigated not only to understand cells from a physicochemical perspective (1, 2) but also to explore diverse applications, such as the fabrication of pharmaceutical capsules (3) and soft robotics (4). Typical examples are liposomes and water-in-oil emulsions covered with lipid membranes. By reproducing various biological phenomena, such as biochemical reactions, in artificial cells, we can bridge the gap between living cells and conventional in vitro systems and derive overhead effects that distinguish the two, such as cell size confinement effects (5, 6). In recent years, as liquid-liquid phase separation (LLPS) in living cells has attracted attention, there has been a growing movement to use condensed droplets formed by LLPS as artificial cells (7-10). LLPS droplets can transport macromolecules to and from the outside, which is not possible with artificial cells covered with semipermeable lipid membranes.

One of the most typical aqueous two-phase systems (ATPS) that facilitates the formation of LLPS droplets is binary polymer mixtures of polyethylene glycol (PEG) and dextran. Previous reports demonstrate that dextran-rich droplets selectively incorporate various macromolecules from the outside, such as long DNA (e.g., lambda DNA with contour length > 15 μm) and actin filaments (length >> 1 μm) (7-9). These macromolecules can be trapped inside the LLPS droplets for a prolonged duration by adding small liposomes, where the liposomes coat the LLPS droplets and stabilize them (8, 11, 12). Recently, PEG concentration gradients generated by local water addition or evaporation enabled dextran droplets to undergo a propelling motion (13, 14). It is strongly suggested that the driving force for the propelling droplet is Marangoni convection caused by the PEG concentration gradient around the droplet. Such motile LLPS droplets through the Marangoni convection are reported for bovine serum albumin (BSA) droplets in PEG with chemical gradients (15, 16). However, the gradient around the motile LLPS droplets has not been quantitatively evaluated, and its direct correlation with convection inside and outside the droplet, motility, molecular transport, *etc.* is unknown.

In this study, we placed the dextran droplets in an elongated chamber with a constant PEG concentration gradient. By analyzing the polymer composition inside and outside of the droplets, we revealed that local PEG concentration gradients lead to Marangoni convection and drive the droplets to move. Furthermore, we found that long DNA was spontaneously incorporated into dextran droplets, and DNA conformation altered its distribution inside the dextran droplets and the motile velocity. These findings demonstrate that the coupling between Marangoni convection and phase separation among the different polymers can determine the structure of the LLPS droplets (*i.e.*, molecular arrangement inside and outside the droplets) and the motile speed.

## Results

### Motile dextran droplet moving down a PEG concentration gradient

In this study, we used a binary blend of 8 wt% PEG and 6 wt% dextran to prepare dextran droplets in PEG solution (Figure 1a, left; see also S1 in the Supporting Information (SI)). For fluorescence observation, 0.1 mol% of Rhodamine-B-labeled PEG (RB-PEG, red) and fluorescein-isothiocyanate-labeled dextran (FITC-dextran, green) were added to the original polymer mixtures. Dextran droplets were prepared by taking both the PEG-rich upper phase and the dextran-rich lower phase and remixing them in a volume ratio of 18:0.6 for 1 minute with a vortex mixer (Figure 1a, right). Under the experimental condition, the initial compositions estimated from fluorescence intensity analysis are 10.5% PEG and 0.44% dextran for the PEG-rich phase, and 3.6% PEG and 20.0% dextran for the dextran droplet. This initial composition is indicated by the green diamond in the phase diagram (Figure 1a, left). The PEG solution containing the dextran droplets is poured into the elongated chamber (Figure 1b). To the solution, 1/3 volume of water is added from one side to provide a PEG concentration gradient. The $x$-axis is set along the long side of the chamber with its origin at the end opposite where water was added.

To estimate the PEG concentration $C_p$ profiles along the $x$-axis and the time course, we analyze the local fluorescence intensity of RB-PEG after the water addition for a PEG-only system with initial $C_p$ = 10 wt% (see S2 in SI and Figures S1, S2). The figure 1c shows the time course of $C_p$ along the $x$-axis (left) and at a position far from the water addition point ($x$ = 5 mm) and at a position close to it ($x$ = 11 mm) (right). It was found that the $C_p$ values for $x \leq$ 6 mm and for $x \geq$ 12 mm maintained the initial $C_p$ (~10 wt%) or <1/5 of the initial $C_p$ (<2 wt%) from just after the water addition until 900 sec later. This means that when water is poured from one side between the thin glass plates by capillary force, the water immediately mixes with PEG at $x \geq$ 12 mm, but not at $x \leq$ 6 mm, resulting in a constant $C_p$ gradient for 6 mm < $x$ < 12 mm. The $dC_p/dx$ for 6 mm < $x \leq$ 7 mm is ~3.0 wt% mm$^{-1}$, which is larger than that of ~0.9 wt% mm$^{-1}$ for 7 mm < $x$ < 12 mm (see triangles in Figure 1c, left). Since the diffusion coefficient of PEG in a 5-20 wt% PEG solution is 20-50 μm$^2$/sec (17), the thin chamber geometry or the interaction between the solution and glass surface may contribute the constant $C_p$ gradient formation for 6 mm < $x$ < 12 mm.

After the addition of water, the dextran droplets at 6 mm < $x$ < 12 mm begin to move in a straight line toward the lower concentration of PEG (*i.e.* the increasing direction of the $x$-axis). Figure 1d shows the snapshots for 30 sec (see Movie S1), where the color bars indicate the time course of the droplet movement. This kind of movement was not observed for droplets before the addition of water (Figure S3). These data strongly suggest that the $C_p$ gradient along the $x$-axis is the driving force for the droplet motion.

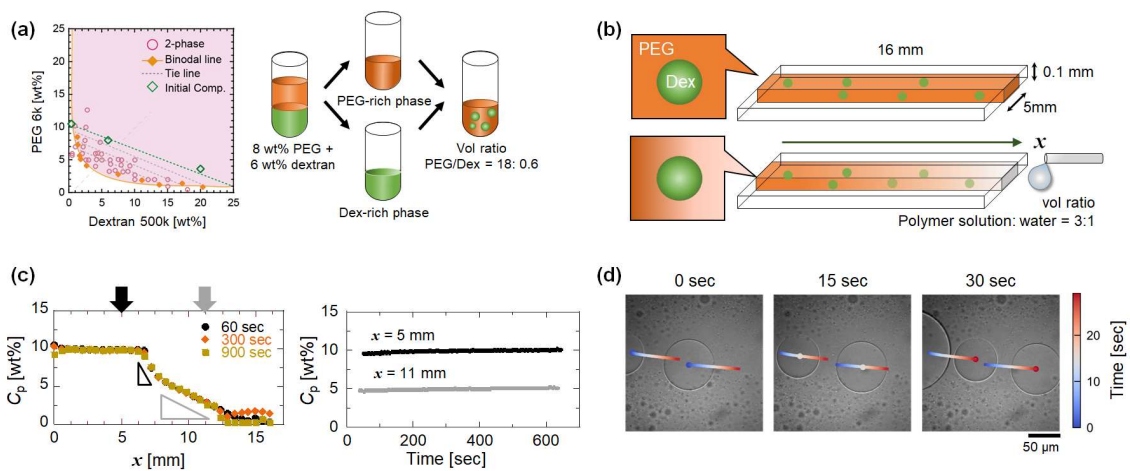

**Figure 1.** Dextran droplets moving down a concentration gradient of PEG solution. (a) Schematic illustration of the phase diagram of PEG and dextran mixtures in bulk at 20 °C and experimental setup. 2- and 1-phase regions are shown in pink (circle) and white, respectively. The binodal line and tie lines are indicated by a solid orange curve and dotted gray lines. (b) Schematic of the elongated chamber for dextran droplets in the PEG-rich phase. Adding water from one side provides a gradient of PEG concentration $C_p$ along the $x$-axis. (c) (left) Time course of $C_p$ gradient along the $x$-axis at a height of ~20 μm after the addition of water for PEG solution with initial $C_p$ = 10 wt%. The triangles suggest the magnitude of the local $C_p$ gradient. (right) Time course of $C_p$ at $x$ = 5, 11 mm (indicated as arrows in the left figure) (d) Snapshots of dextran droplets moving down $C_p$ gradient (see movie S1). The droplet trajectory for 30 sec and the direction of movement are shown by the solid line. The phase diagram in panel (a) is reprinted with permission from Ref. (18) Copyright 2022 American Chemical Society.

To quantify the $C_P$ gradient around the motile droplets, here we estimate the local $C_P$ gradient along the $x$-axis, $dC_P/dx$, from the fluorescence intensity of RB-PEG as follow:

$$\frac{dC_P}{dx} \propto \frac{|I_f - I_b|}{|x_f - x_b|}, \tag{1}$$

where $I_f$ and $I_b$ are the average fluorescence intensities at the motile droplet front $x_f$, and the back $x_b$, respectively (Figures 2a, 2b, 2c; see also S2, S3 in SI and Figures S2). To simplify the analysis, we assumed that the intensity difference and the concentration difference between the two points are linearly proportional to the length $(x_f - x_b)$. The length $(x_f - x_b)$ was set to 100 to 150 μm, which is sufficiently shorter than the 16 mm long side of the chamber (Figure 1c). Figure 2d clearly shows there is a finite PEG concentration difference in front and back of the droplet. The values of $dC_P/dx$ are somewhat scattered, but $dC_P/dx$ appears to decrease stepwise with time, which is confirmed by the histograms having two peaks (see the inset in Figure 2d). The average value before and after 300 sec is 0.032 mg/mL μm$^{-1}$ (~3 wt% mm$^{-1}$) and 0.015 mg/mL μm$^{-1}$ (~1.4 wt% mm$^{-1}$), respectively. These are comparable to the values obtained with PEG only system (approximately 3 and 0.9 wt% mm$^{-1}$ for 6 mm $< x \leq$ 7 mm and 7 mm $< x <$ 12 mm, as shown in Figure 1c). This means that after ~300 sec the droplets passed through the large $dC_P/dx$ region 6 mm $< x \leq$ 7 mm and reached the small $dC_P/dx$ region at $x >$ 7 mm.

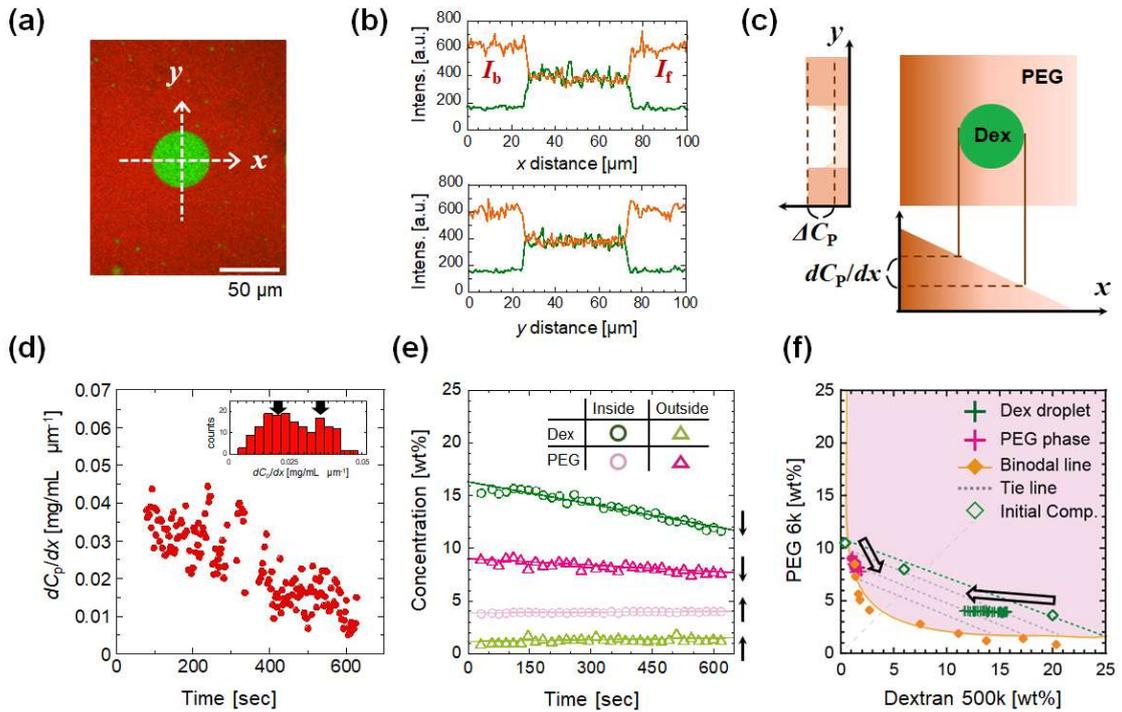

**Figure 2**. Compositional shifts associated with droplet motion. (a) Confocal fluorescent image of a dextran droplet (green) in a PEG-rich phase (red). (b) Fluorescence intensity profiles along the *x*-axis (direction of PEG concentration gradient, upper) and the perpendicular *y*-axis (lower) along the *x*-axis shown in (a). (c) Schematic illustration of the PEG concentration difference inside and outside of the droplet $\Delta C_P$ and the gradient along the *x*-axis. (d) Local PEG concentration gradient $dC_P/dx$ between the front and back of the droplet versus time after water addition. The inset shows a histogram of $dC_P/dx$ values, showing two peaks indicated by allows. (e) Time course of the concentrations of PEG (pink) and dextran (green) inside (circles) and outside (triangle) the dextran droplets. (f) The compositional shifts shown in (e) are mapped onto the phase diagram. The arrows show the time course.

Accordingly, we analyzed the time development of compositional changes inside and outside of the motile droplets at the same *x*-position from the fluorescence intensity profiles along the *y*-axis (Figure 2b). Figure 2e shows the time course of PEG concentration $C_P$ (pink) and dextran concentration $C_D$ (green) for inside the dextran droplets (circles) and the outside (triangles). Inside the dextran droplet, $C_D$ decreased, and $C_P$ increased with time. In contrast, outside the dextran droplet, $C_D$ increased and $C_P$ decreased. In other words, both inside and outside the droplet, the concentrations of the major components decreased, and the minor components increased. Mapping these compositional changes over time on a phase diagram clearly shows that the motile droplet repeatedly takes in and out its constituent polymers (Figure 2f). As a result of this compositional change, the average interfacial tension at the droplet surface would be reduced with time by a factor of about 50 (Figure S4a). However, $C_p$ gradient in the chamber should result in a finite interfacial tension difference in front of and behind the droplet and Marangoni convection around the droplets.

After several tens of minutes, some droplets coalesce and grow. Large droplets with a radius $R > 50$ μm adhered to the glass surface and stopped moving. However, convections inside and outside of the droplets maintained (see Movie S2) as shown in the 120 sec trajectories of PEG domains (black) inside and the small dextran domains (green) outside (see arrows in Figure 3a). These trajectories show the convections both inside and outside the droplet surface, as illustrated in Figure 3b. These data support our idea that the difference in interfacial tension caused by $C_p$ gradient causes Marangoni convection and droplet motion, as suggested previously (13, 14). We analyzed the convections in detail based on the squirmer model in a Brinkman medium (19-21) and the fitting are shown as the blue arrows in Figure 3a (see also S4 in SI for the analysis) (22). It suggests that the convection outside the droplets is dipolar ($\beta = -0.1$, nearly neutral). The speed of the convections is high near the droplet boundary and the center. Marangoni convection has traditionally been visualized by the addition of guest particles (23, 24), but in the present system, isolated minor-phase domains (PEG domains in dextran droplets and dextran domains in the PEG-rich phase) make the convections visible.

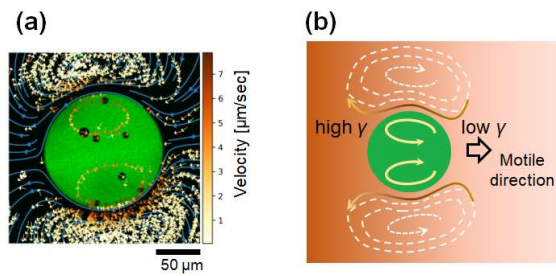

**Figure 3.** Convection inside and outside the large droplet in $C_p$ gradient ($C_p$ is lower on the right side). (a) The 120 s trajectories of small dextran domains (green) and small PEG domains (black) outside and inside the droplet. The color indicates that the magnitude of the velocity. Blue arrows are fitting lines using the squirmer model of the Brinkman medium ($\beta = -0.1$, nearly neutral) (see S4 in SI). (b) Schematic illustration of the convection inside and outside the dextran droplet (green) in the concentration gradient of PEG (orange).

**Motile dextran droplet with DNA**

For the phase-separated solution of PEG and dextran, long DNA is reported to preferentially localize in the dextran-rich phase (7, 12, 25). To prepare the motile dextran droplets containing DNA, here we added 0.32 mg/mL coil DNA (lambda DNA with 48 kbp) in 0.3 mM Tris-HCl (see also S1). Fluorescence cross-sectional images (Figure 4a, left), 3D images (Figure S5), and reconstructed images from the radial profile (Figure S6) show that the droplets with DNA maintain their almost spherical shape. Fluorescence intensities along the *z*-axis of droplets (Figure S5) demonstrates that the coil DNA molecules are homogeneously distributed inside the droplets before the addition of water. After the addition of water, the dextran droplets start to move toward a lower PEG concentration, while holding the DNA (Figure 4a, right). After a short time, the rear of the moving droplet appears black, and several black lines extending from it are observed. It means that coil DNA moves away from the central axis which separates the convections (Figure 4b). Covering the rear of the droplet with small liposomes (shown in yellow in Figure 4c) vanished the black lines at the droplet center and increased the triangle zone at the rear. It suggests that coating the rear of motile droplets with liposomes may allow control of DNA distribution within the droplets.

To investigate the effect of the coil-globule transition of DNA on the distribution inside the droplet and the droplet motion, 20 mM MgCl$_2$ was added to the droplets containing coil DNA. The globule DNA particles appear to be numerous bright spots inside the droplet (Figure 5a, left edge). When convections appear inside the droplet after the addition of water, the globule DNA particles are transported towards the droplet's rear along the droplet surface via Marangoni convection, forming a large aggregate (Figures 5a, 5b). This localization and aggregation of globule DNA molecules at the droplet rear part was not observed for coil DNA (Figure 4a).

Unfortunately, we cannot analyze the convections around the motile droplets because the small dextran domains were less observed compared to systems without DNA. It should be noted that the fusion process between two droplets with coil DNA takes more than 3 sec (Figure 4d). This is much longer than the < 1 sec for droplets without DNA (see Movie S1) and for droplets with globule DNA (Figure 5c). This slower coalescence process of coil DNA than of globule DNA suggests that the droplets with coil DNA have a higher dynamic viscosity than those with globule DNA.

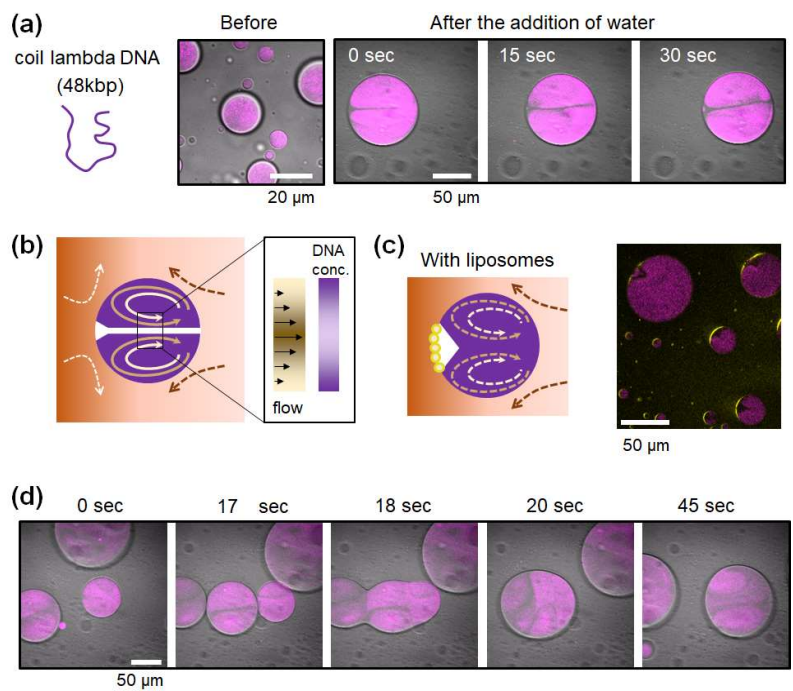

**Figure 4**. Motile dextran droplet containing lambda DNA in coil (purple). (a) Time series of droplets before and after the addition of water. The average velocity $v$ is 2.34 μm/sec. (b) The schematic of possible convection inside and outside the droplet. The enlarged image shows the expected DNA concentration under the flow at the center of the droplet. (c) Fluorescence image of motile droplets covered with small liposomes (yellow) at the rear and the illustration of possible convections. (d) Coalescence of motile dextran droplets moving in the same direction.

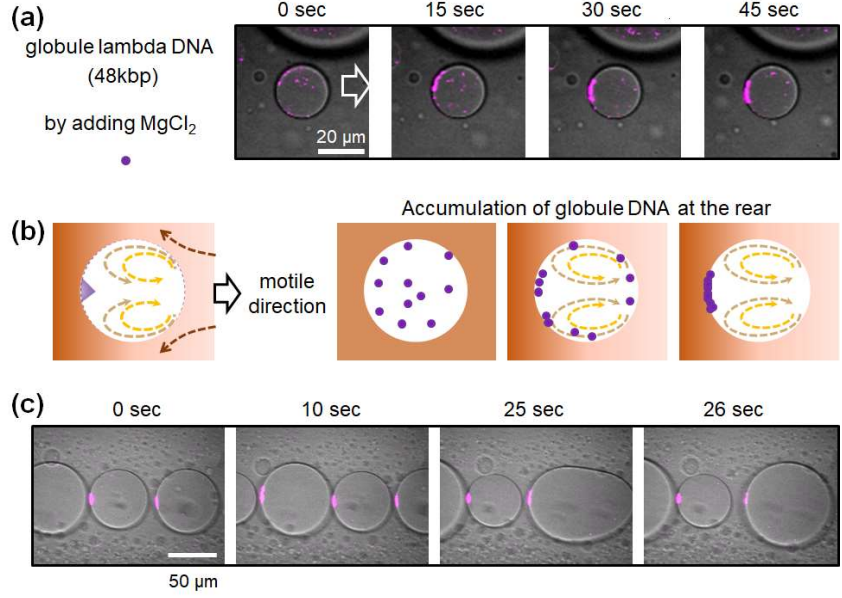

**Figure 5**. Motile dextran droplet containing globule DNA (purple) in the presence of 20 mM MgCl$_2$. (a) Time series of globule DNA inside the droplet after the addition of Mg$^{2+}$ solution, and (b) the schematic of possible convection inside and outside the droplet. (c) Coalescence of motile dextran droplets moving in the same direction.

**Motile speed of dextran droplet with or without DNA**

To clarify the DNA's conformation effect on the droplet motile velocity $v$, here we analyze the position of the droplet center of gravity with time. Only spherical droplets with a radius of 50 μm or less were used for velocity analysis to eliminate the influence of droplet deformation due to glass adhesion. The velocity $v$ of a droplet of radius $r$ is faster for larger droplets, as reported previously (13). In addition, the normalized velocity $v/r$ becomes constant at 200−300 sec after the addition of water, irrespective $r$ (Figure 5a). This 200−300 sec corresponds approximately to the time it takes for a droplet to pass through the large $dC_P/dx$ region near $x = 7$ mm (Figures 1c, 2d).

Ces et al. reported that $v/r$ follows as (13) (similar equations are shown in (26) and Ref. 41-48 in (13))

$$\frac{v}{r} = \frac{\partial C_p}{\partial x} \frac{\partial \gamma}{\partial C_p} \frac{1}{2\eta_o + 3\eta_i}, \tag{2}$$

where $C_p$ is the PEG concentration outside the droplet, $x$ is the position of the droplet center of gravity, $\gamma$ is the interfacial tension between the dextran-rich phase and the PEG rich phase, and $\eta_o$ (or $\eta_i$) is the solution viscosity outside (or inside) the droplet. In our experimental system, Reynolds number Re is estimated to be <<1 (for example, Re = $\rho v r/\eta_o \sim 10^{-6}$ by using $\rho = 1.01 \times$ $10^3$ kg/m$^3$, $v$ = 1 μm/sec, $r$ = 10 μm, $\eta_o$ = 10 mPa sec) so that the contributions of Marangoni convection and drag forces to the velocity should balance immediately. To verify the applicability of Eq. (2) to our dextran droplet, normalized velocity $v/r$ for droplets with $r$ = 8−46 μm was analyzed for four different compositions: PEG/dextran system as a control (white), PEG/dextran with DNA in coil (pink) and in globule by adding 20 mM MgCl$_2$ (green), and PEG/dextran with 20 mM MgCl$_2$ (gray) (Figure 6b). The value of $v/r$ of PEG/dextran with coil DNA is 0.06 sec$^{-1}$, approximately half the value of other systems, ~0.1 sec$^{-1}$.

To explain the smaller $v/r$ of dextran droplets with coil DNA, we measured the interfacial tension between PEG-rich and dextran-rich phases (Figure 6c) and dynamic viscosity for dextran-rich phase (Figure 6d). The interfacial tension does not change much with the presence or absence of DNA or ions (Figures 6c, S4b). On the other hand, the dynamic viscosity at a low shear rate of <10$^0$ sec$^{-1}$ for dextran-rich phase with coil DNA was >2 times higher than the others (Figure 6d). Therefore, we concluded that the smaller $v/r$ of the coil DNA system than that of the other systems is due to the increase in the dynamic viscosity inside the dextran droplet with coil DNA, rather than a difference in interfacial tension gradient. This is consistent with the flow and fusion process being slower with the coil DNA system than the other systems (Figures 3, 4d, 5c, S7). In addition, the dextran-rich phase containing coil DNA exhibited shear thinning as observed in the dense DNA solution (27). This can be interpreted as follows: at low shear, the viscosity is high because dextran entangles with the coiled DNA strands, whereas at high shear, the DNA strands are stretched, which disentangles them and reduces the viscosity to the same level as the other systems. It should be mentioned that the diffusion coefficient of dextran chains in a droplet under no convective conditions is independent of the presence of DNA or ions (Figure S9).

In the case of PEG/dextran system, the estimated $v/r$ based on Eq. (2) is ~0.1 sec$^{-1}$ by applying $dC_p/dx$ = 0.02 mg/mL μm (Figure 2d), $d\gamma/dC_p$ = (0.06 mN/m) / (20 mg/mL), and $\eta_o$ = 3 mPa sec (28), and $\eta_i$ = 200 mPa sec. This estimated $v/r$ is consistent with the experimental values except the system with coil DNA (Figure 6b).

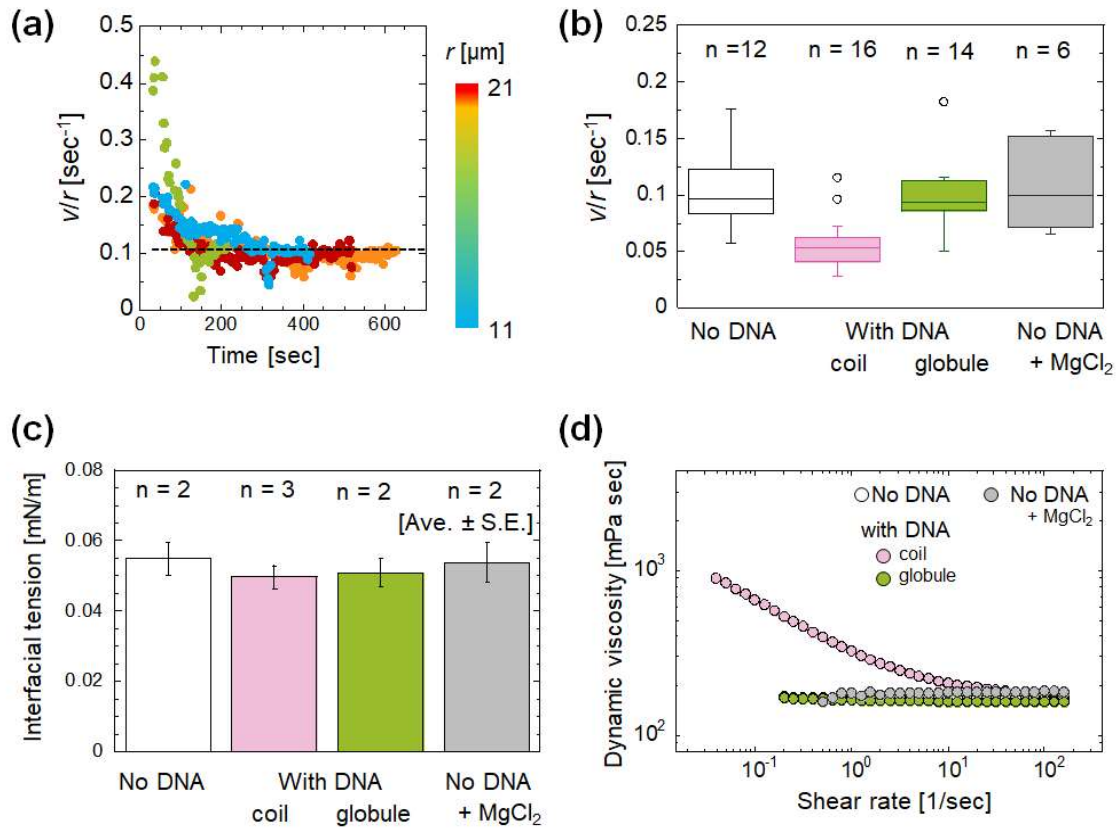

**Figure 6**. Droplet motile speed and the driving and restraining forces. (a) Time development of the normalized velocity $v/r$ of the motile dextran droplets with a radius $r$ after the addition of water. The color shows $r$ at 200 sec. For example, the average velocity $v$ is 1.7, 2.5, 2.7, and 2.0 μm/sec for a droplet with $r$ = 11, 14, 18, and 21 μm, respectively. (b) Composition dependence of $v/r$ for dextran droplets (white), dextran with DNA in coil (pink) and in globule by the addition of 20 mM $MgCl_2$ (green), and dextran droplets with 20 mM $MgCl_2$ without DNA (gray). (c) Interfacial tension between PEG-rich and dextran-rich phases (See also Figure S4 for control experiments). (d) Dynamic viscosity vs. shear rate for only dextran-rich phases (white), dextran-rich phases with DNA in coil (pink) and in globule by the addition of 20 mM $MgCl_2$ (green), and dextran-rich phases with 20 mM $MgCl_2$ only (gray).

**Discussion**

We quantitatively demonstrated that Dextran droplets move due to Marangoni convection driven by a constant concentration gradient of PEG formed in an elongated chamber (Figures 1, 2, 3). The droplet speed normalized by the droplet's radius was explained by the local PEG concentration gradient around the droplet, the corresponding interfacial tension difference, and dynamic viscosity (Figure 6). We confirmed that the final composition of the PEG-rich phase is almost identical to the thermal equilibrium phase after the water addition (7.3 wt% PEG and 1.4 wt% dextran). On the other hand, the dextran-rich phase deviates from the thermal equilibrium phase, which may be related to the slower diffusion of dextran ~0.2 $\mu m^2$/sec (Figure S9) compared to that of PEG outside, 20-50 $\mu m^2$/sec (17).

According to the coil-globule transition of DNA, DNA changes its position inside the motile dextran droplets. For the coil DNA, the DNA was eliminated from the center of droplets with convections along the motile direction, as shown in the black center line (Figure 4a). This could be due to shear banding of DNA caused by convective gradients at the center of the droplet (Figure 4b), since the dextran solution with coil DNA exhibits shear thinning behavior (Figure 6d). Inflow of PEG solution can also contribute to the DNA's elimination from the center line. On the other hand, globule DNA was accumulated at the rear parts under the convection (Figure 5a). It is reported that the aggregation of globule DNA is proceed via depletion interaction by a crowder molecule such as PEG and electrostatic interaction by cations like $Mg^{2+}$ (29). In addition, globule DNA adheres to the dextran droplet surface if it contact by diffusion without the convection (25). Therefore, we concluded that the observed localization and aggregation of globule DNA at the rear is caused by the coupling of DNA conformation, Marangoni convection, and the DNA adhesion to the droplet surface. The zeta potentials of the droplet surface with DNA will help us to discuss the mechanism of the DNA localization in more detail. In addition, this localization can be altered by coating the motile droplet surface with liposomes (Figure 4c) or colloids (30).

Like the dextran droplets without DNA, we try to track the isolated small domains around the motile droplets with coil DNA and fitted them based on the squirmer model (see Figure S7). As shown by the warm-color arrows, the averaged flow velocity for the coil DNA system is slower than droplet without the DNA (Figure 3a). It also supports that the dynamics viscosity inside the dextran is higher for DNA coil system than the other systems (Figure 6d). In addition, their fitting lines, shown in blue arrows, suggests that the convection outside the droplets is quadrupolar ($\beta$ = 2.5, puller) for coil DNA system, unlike the dipolar ($\beta$ = −0.1, nearly neutral) for without DNA system. Therefore, the convections might be altered by coexisting molecule and its conformation.

To consider the effect of convection on the multi-droplet motion, here we simulate the time development of two droplets with the same radius $r$ moving on the same line (see S5 in SI for the simulation condition). In our experimental system, Reynolds number Re (= $\rho v r / \eta_o$ ~ $10^{-6}$ by using $\rho$ = 1.01 × 10$^3$ kg/m$^3$, $v$ = 1 μm/sec, $r$ = 10 μm, $\eta_\mathrm{o}$ = 10 mPa sec) is small enough that inertial effects are negligible, resulting in linear and laminar Stokes flow. We set the initial inter-droplet distance $d$ to be 2$r$. In the case of a neutral type ($\beta$ = 0), the droplets move with the velocity, $v$ and do not collide (Figure S10a). On the other hand, puller type droplets ($\beta$ = 0.1) approach together due to the attractive force due to the intake flow behind the droplet (Figure S10b). The change in $d$ with time slows down just before the collision (Figure S10c). This is mainly due to the lubricating force of the fluid between the puller type droplets (31). These results suggest that the coexisting molecules inside the motile droplets not only alters the motile speed of individual droplets, but also the motile behavior of multiple droplets, similar to the oil droplets and solid spheres covered by surfactant membranes (32-36). In addition to such a long-range attractive force between puller type droplets (19-22), there may be short-range repulsive forces between DNA-containing droplets due to the negative surface charges (37). These complex short- and long-range interactions between motile droplets are thought to determine the process by which droplets approach each other and eventually coalesce (Figures 4d, 5c).

**Summary and Conclusion**

In this paper, we clarified the qualitative relationship between the motility and concentration gradient for Marangoni droplets of dextran in PEG. The motility is altered by coexisting DNA sensing the presence of Mg$^{2+}$. The coil-globule transition of DNA accelerates the droplet's motile speed due to the decrease in droplet's dynamic viscosity and the spatial distribution of DNA inside the droplets. Combining this system with chemically-driven Marangoni convections (38-40), more complex movement for motile droplets will be expected. Incorporation of various biopolymers should result in shear-induced segregation (41) and also the alignments of stiff polymers such as collagen fibrils by the flow (42). We expect that our findings about the active flow-induced segregation of DNA inside the motile droplets will contribute to the comprehension of self-organization mechanisms through phase separation under convection and to application research by using them as soft robotics (43).

**Notes**

The authors declare no competing financial interest.

**Table of Contents graphic**

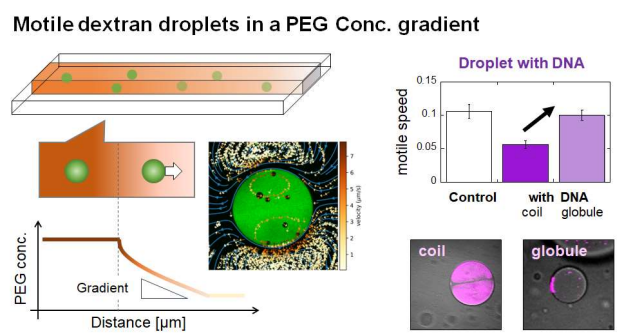

**Supporting Information**

Materials and methods (S1), Estimation of the concentration gradient of PEG (S2), Estimation of concentration difference inside and outside of the droplet (S3), Flow analysis by tracking the small domains around the droplet (S4), Simulation of two motile droplets (S5). Linear relationship between the RB-PEG fluorescence intensity and PEG concentration (Figure S1), fluorescence image for estimating the concentration gradient of PEG (Figure S2), the Time series of dextran droplets without PEG concentration gradient (Figure S3), the Comparison of interfacial tensions (Figure S4), the 3D image of droplets containing DNA (Figure S5), and the shape of motile droplets (Figure S6), convection analysis for droplets with coil DNA (Figure S7), the FRAP experiment for dextran droplets after the whole bleaching (Figure S8), the Diffusion coefficient of dextran inside dextran droplets (Figure S9), the Simulation data for two motile droplets and the inter-droplet distance (Figure S10). Movie of a motile dextran droplet (Movie S1), movie of a small PEG-rich domain within a large dextran droplet (Movie S2).

**Acknowledgement**

We thank Prof. Masatoshi Ichikawa at Kyoto University and Prof. Natsuhiko Yoshinaga at Tohoku University for useful discussions. This research was funded by the Japan Society for the Promotion of Science (JSPS) KAKENHI (grant numbers 22H01188, 24H02287 (M.Y.) and 21K15057 (H.S.)) Japan Science and Technology Agency (JST) Program FOREST (JPMJFR213Y (M.Y.)).

# Supporting Information for

# Marangoni droplets of dextran in PEG solution and its motile change due to coil-globule transition of coexisting DNA

Tomohiro Furuki[1,2,†], Hiroki Sakuta[3,4,†], Naoya Yanagisawa[3], Shingo Tabuchi[5], Akari Kamo[5], Daisuke S. Shimamoto[3], Miho Yanagisawa[2,3,4,5*]

[1] Faculty of Pure and Applied Sciences, University of Tsukuba, Tennodai 1-1-1, Tsukuba, Ibaraki 305-8573, Japan

[2] Department of Integrated Sciences, College of Arts and Science, The University of Tokyo, Komaba 3-8-1, Meguro, Tokyo 153-8902, Japan

[3] Komaba Institute for Science, Graduate School of Arts and Sciences, The University of Tokyo, Komaba 3-8-1, Meguro, Tokyo 153-8902, Japan

[4] Center for Complex Systems Biology, Universal Biology Institute, The University of Tokyo, Komaba 3-8-1, Meguro, Tokyo 153-8902, Japan

[5] Graduate School of Science, The University of Tokyo, Hongo 7-3-1, Bunkyo, Tokyo 113-0033, Japan

**AUTHOR INFORMATION**
*Corresponding Author: myanagisawa@g.ecc.u-tokyo.ac.jp

**Author contributions**: T.F. and H.S. equally contributed to this research.
M.Y. designed research; T. F., H. S., N. Y. and A. K. performed experiments and T. F., H. S., and M. Y. analyzed the data; S. T. and D. S. performed simulation and analyzed the data. All authors wrote the paper.

**Contents**

## S1. Materials and methods

### a. Materials

To prepare binary polymer blends of polyethylene glycol (PEG) and dextran, we used dextran from *Leuconostoc spp.* with a molar mass of 450–650 kg/mol (dextran 500k; Ca# 31392; Sigma-Aldrich St. Louis, MO, USA), PEG with a molar mass of 7–9 kg/mol (PEG 6k; Ca# 169-09125; FUJIFILM Wako Pure Chemical Co., Japan; Tokyo, Japan), and ultrapure distilled water (Invitrogen, CA, USA). As a fluorescent-labeled polymer, we used fluorescein-isothiocyanate-labeled dextran with a molar mass of 500 kg/mol (FITC-dextran, Sigma-Aldrich) and rhodamine-B-labeled PEG with a molar mass of 5 kg/mol (RB-PEG; Nanocs, MA, USA). Additionally, Lambda DNA, SYBR™ Green I Nucleic Acid Gel Stain, $MgCl_2$, and Tris-HCl (pH 8.0) were purchased from Nippon Gene (Tokyo, Japan), Takara Bio (Shiga, Japan), Wako Pure Chemical Co., Japan (Tokyo, Japan), and Nacalai Tesque (Kyoto, Japan), respectively. 1,2-dipalmitoyl-sn-glycero-3-phosphocholine (PC) and 1,2-dimyristoyl-sn-glycero-3-phosphoethanolamine-N-(lissamine rhodamine B sulfonyl) (ammonium salt) (RB-PE) were purchased from Avanti Polar Lipids Inc. (AL, USA).

### b. Preparation of dextran droplets in PEG-rich phase

Stock solutions of PEG and dextran with a concentration of 15–20 wt% were prepared by dissolving them in distilled water at room temperature. After allowing them to stand for 1 day at 4°C, the stock solutions were mixed to achieve 8 wt% PEG and 6 wt% dextran. For fluorescence observation of PEG and dextran phases, approximately 0.1 wt% RB-PEG and 0.1 wt% FITC-dextran were added to the PEG and dextran solutions, respectively. To achieve equilibrium, the PEG and dextran solutions were vortexed for 1 min followed by centrifugation (5000 rpm, room temperature, 1 hour). Each concentration of PEG and dextran in the upper and lower phases was determined by measuring the refractive index and density for each phase as previously reported (1), and was 10.5 wt% PEG and 0.44 wt% dextran for the upper PEG-rich phase and 3.6 wt% PEG and 20 wt% dextran for the lower dextran-rich phase. For the lambda DNA experiments, lambda DNA was mixed in a 10 mM Tris-HCl (pH 8.0) solution with 1xSYBR Green instead of FITC-dextran, which has nearly the same fluorescence wavelength range. To obtain dextran-rich droplets with or without 0.32 mg/ml DNA, 18 μL upper PEG-rich phase, 0.6 μL lower dextran-rich phase, and 1.4 μL DNA solution were mixed and vortexed for 1 minute to obtain droplets (Figure 1a). Then, 6 μL of the droplet solution was poured into an elongated chamber that was 5 mm wide, 16 mm long, and 0.1 mm deep (Figure 1b, upper). The chamber consisted of two pieces of cover glasses (Matsunami Glass Ind., Ltd., Osaka, Japan) attached with double-sided adhesive tape. To initiate the coil-globule transition of lambda DNA, 20 mM $MgCl_2$ was added to the above PEG-dextran solution with DNA. To provide a PEG concentration gradient, 1/3 volume (2 μL) of water was added from one side of the chamber (Figure 1b, lower).

**c. Preparation of small liposomes**

10 mM PC in chloroform with 0.4 mol% RB-PE was dried under nitrogen flow and stored in vacuum overnight to obtain a lipid film. 1.0 mL of PEG-rich phase mixed with lipid film was sonicated for 30 min to obtain a liposome solution (~7 mg/mL). Then, the liposome solution was extruded with 21 passes with a 0.1 μm pore filter (Whatman filters, Avanti Mini-Extruder) to produce a liposome solution with a uniform diameter. This liposome solution was added to the PEG/dextran solution at 1:2 (vol/vol).

**d. Fluorescence observation of motile dextran droplets**

Motile droplets were observed using a confocal laser scanning microscope (IX83, FV1200; Olympus Inc., Tokyo, Japan) equipped with a water immersion objective lens (UPLSAPO 60XW, Olympus Inc.). FITC-dextran (or SYBR Green) and RB-PEG (or RB-PE) were excited at wavelengths of 473 nm and 559 nm and detected in the ranges of 490–540 and 575–675 nm, respectively. Movies of the motile droplets were obtained at 1 fps. The fluorescence intensity was analyzed using Fiji software (National Institute of Health (NIH), USA). In addition, we used manual fitting for fluorescence images of dextran and the TrackMate plugin in Fiji software (National Institute of Health, USA) to analyze motile velocity and droplet radius (2).

**e. Interfacial tension measurements**

Interfacial tension between the PEG-rich phase and the dextran-rich phase was measured by a spinning drop tensiometer (SDT, Krüss, Hamburg, Germany). Four different compositions were used for the measurements: (i) 6 wt% PEG and 6 wt% dextran solution, (ii) containing 20 mM $MgCl_2$, (iii) containing 0.32 mg/ml DNA, (iv) containing 0.32 mg/ml DNA and 20 mM $MgCl_2$. The DNA and $MgCl_2$ in 0.25 mM Tris-HCl (pH 8.0) were added to the dextran-rich solution, as mentioned above. These solutions were vortexed for 1 min, followed by overnight incubation and centrifugation (5000 rpm, room temperature, 30 min). 50 μL PEG-rich solution and 1 mL dextran-rich solution were used for each measurement at 25–27.5 °C.

**f. Static viscosity measurements**

Dextran diffusion analysis using FRAP (Fluorescence Recovery After Photobleaching) was used to estimate the viscosity of the dextran droplets before the PEG concentration gradient. FRAP experiments were performed as described elsewhere (3, 4). All samples contained FITC-dextran but no other fluorescent reagent. At the center of droplets with a radius of 15–25 μm, circular regions with a radius of 4 μm were breached for 1 sec with 100% laser power. The intensity of 5 sec before and 175 sec after breaching were obtained at 1 fps. We derived the diffusion coefficients $D$ using a molecular diffusion package (Olympus software). Then, the viscosity $\eta$ of the dextran-rich phase was estimated based on the Stokes–Einstein equation.

$$D = \frac{k_\text{B} T}{6\pi\eta a} \tag{S1}$$

where $k_\text{B}$ is Boltzmann's constant ($1.38\times 10^{-23}$ J /K), $T$ is temperature (298 K), and $a$ is radius of gyration of dextran 500k (approximately, 17.2 nm (5)).

g. **Dynamic viscosity measurements**

Dynamic viscosity [mPa·sec] of dextran-rich phase with or without DNA, and $MgCl_2$ (see S1, b for the sample preparation) was measured by using a rheometer (MCR502, Anton Paar). The viscosity was measured as a function of shear rate by a cone-plate measuring system of 50 mm diameter with the cone angle 1° (CP50-1). For the dextran-rich phase with $MgCl_2$, the solution became cloudy under shear, thus we used a parallel-plate measuring system of 43 mm diameter (PP43/GL). All measurements were done at 25ºC.

**S2. Estimation of PEG concentration and the gradient**

We estimated the PEG concentration from the fluorescence intensity of the RB-PEG, setting the background intensity to zero. The average fluorescence intensity per unit volume was measured using confocal laser fluorescence microscopy while repeatedly diluting 10 wt% PEG with 0.1 wt% RB-PEG. Figure S1 shows that the linear relationship between the PEG concentration $C_\text{p}$ and the fluorescence intensity of the RB-PEG $I$, independent on the laser power. Since this linear relationship holds for the 10 wt% or less PEG solution used in the experiment, we estimated the local PEG concentration from the RB-PEG intensity. To estimate $C_\text{p}$ gradient within the chamber, here we set the $x$-axis as the direction of the long axis of the chamber where the concentration gradient of PEG is created (the side with the highest concentration is origin). We obtained the PEG concentration gradient along the $x$-axis and its time course by moving the microscope stage at ~1 mm sec$^{-1}$.

The $C_\text{p}$ gradient around the motile dextran droplets is also estimated from the analysis of $I$ (Figure S2). The $y$-axis is the direction of the short axis of the chamber perpendicular to it. The PEG concentration gradient along the $x$-axis, $dC_\text{p}/dx$, is estimated as,

$$\frac{dC_\text{P}}{dx} = C_\text{P}^0 \frac{I_\text{f} - I_\text{b}}{|x_\text{f} - x_\text{b}|}, \tag{S2}$$

where $I_\text{f}$ and $I_\text{b}$ is the average intensity of RB-PEG in a circular region of radius 15.5 μm with centroids in front ($x_\text{f}$, $y_\text{p}$) and behind ($x_\text{b}$, $y_\text{p}$) the droplet (Figure S2, a). $y_\text{p}$ was set approximately 50 μm below the droplet center, and length ($x_\text{f} - x_\text{b}$) was set 100–150 μm longer than the droplet radius $R_d$ (Figure S2, b). $C_P^0$ is the ratio between the initial intensity $I_0$ and the initial PEG concentration at the PEG-rich phase, 105 mg/mL (10 wt%).

### S3. Estimation of concentration difference inside and outside the droplet

The concentration difference inside and outside of the dextran droplets for PEG $\Delta C_p$ and dextran $\Delta C_d$ was estimated from the fluorescence intensities of RB-PEG and FITC-dextran at the same $x_d$ position (Figure S2, b). For larger dextran droplets with a radius >20 μm, we analyzed the average intensities in a circular region with a radius of 15.5 μm inside and outside the droplet, $I_d$ and $I_p$ and derived the concentration differences of $\Delta C_p$ and $\Delta C_d$ as follows:

$$\Delta C_p = C_P^0 |I_d - I_p|, \tag{S3}$$
$$\Delta C_d = C_d^0 |I_d - I_p|. \tag{S4}$$

$C_P^0$ and $C_d^0$ are ratios between the initial intensity $I_0$ and the initial PEG or dextran concentrations. Initial concentrations of PEG and dextran were 10.5 wt% PEG and 0.44 wt% dextran for the PEG-rich phase and 3.6 wt% PEG and 20 wt% dextran for the dextran-rich phase.

### S4. Flow analysis by tracking the small domains around the droplet

For flow analysis, fluorescence microscopy images were taken every second for 120 seconds. To facilitate the analysis, we chose a large droplet with a radius > 50 um, with some of the droplets sticking to the bottom glass and not moving (Figures 3, S7). We tracked the PEG domains, which appear black inside the dextran droplet, and the dextran domains, which appear green outside the droplet. The domain radius is 1-10 μm, which is much smaller than the observed radius of moving dextran droplets. For the tracking, we used TrackMate in Fiji software (2).

To analyze the swimmer type of the droplet, we have fitted the flow structures outside the non-moving droplet (i.e., the swimmer frame) by using a 2D squirmer model in a Brinkman medium: (6)

$$u_\rho = \Sigma_{n=2}^2 b_n \left[ n \left(\frac{a}{\rho}\right)^{n+1} - n \frac{a}{\rho} \frac{K_n\left(\frac{\rho}{\lambda}\right)}{K_n\left(\frac{a}{\lambda}\right)} \right] \cos(n\theta) \tag{S5}$$

$$u_\theta = \Sigma_{n=2}^2 b_n \left[ n \left(\frac{a}{\rho}\right)^{n+1} + \frac{a}{\lambda} \frac{K_n'\left(\frac{\rho}{\lambda}\right)}{K_n\left(\frac{a}{\lambda}\right)} \right] \sin(n\theta) \tag{S6}$$

$(\rho, \theta)$ are polar coordinates with the droplet center as the origin. $a$ is droplet radius and $\lambda = h/2\sqrt{3}$ with $h$ as the depth of the chamber. $K_n$ are the Bessel functions of the second kind. The swimmer type is characterized by $\beta = b_2/b_1$. The fitting parameters were chosen so that the cosine similarity of the flow direction from the model and applicable flow directions obtained from tracking was maximized.

**S5. Simulation of two motile droplets**

To examine the effect of convection on the multi-droplet movement, we simulate the movement of two motile droplets by adopting a 2D squirmer model (7) with the immersed boundary lattice Boltzmann method (8). The squirmer model is implemented by the following boundary condition of the flow velocity at the swimmer boundary in the swimmer frame $\boldsymbol{u}_b$.

$$\boldsymbol{u}_b = B_1(\sin\theta + \beta\sin2\theta)\boldsymbol{e}_\theta \quad (S7)$$

where $\boldsymbol{e}_\theta$ is the tangential unit vector(7). The constant $B_1$ controls the velocity of swimming droplets $v$, while $\beta$ controls the swimming mode. We employ a single relaxation time, nine velocity model (D2Q9) lattice Boltzmann equation to describe the motion of the fluid, while using the same forcing scheme used in (8). For the boundary, we adopt the method in (9), which uses the direct-forcing method. On a boundary point where the velocity is $\boldsymbol{v}_b$, we set the flow boundary condition to be $\boldsymbol{v}_b + \boldsymbol{u}_b$ to impose the boundary condition of the squirmer model, Eq. S5. The magnitude of $\boldsymbol{v}_b$ is the same as the velocity at the center of gravity of the swimming droplet, $v$. The force on the fluid is estimated by interpolating the force applied on the boundary. We use the same interpolation scheme as in (8). The simulation was carried out on a periodic boundary 256 × 256 [lattice units$^2$] sized grid with two identical swimmer droplets with a radius of 5 lattice units, initialized with an inter-droplet distance (surface to surface) of 10 lattice units. The initial motile directions of the droplets are the same, and one droplet is set directly in front of the other. To correspond the numerical units with experimental units, we set one lattice unit length to be 10 μm and one-time step to be 0.001 sec. The speed of the swimmers in the simulations was ~0.0006 lattice units per time step (corresponding to ~6 μm/sec). The time development of the inter-droplet distance $d$ is fitted with the following velocities for the far field and near field, respectively (10).

$$\boldsymbol{v}_{\text{far}} \sim (2r + d)^{-\alpha} \quad (S8)$$

$$\boldsymbol{v}_{\text{near}} \sim \frac{d}{r}\log\frac{d}{r} \quad (S9)$$

The exponents $\alpha$ for quadrupole and dipole cases are 3 and 2, respectively.

**Supplemental Figures**

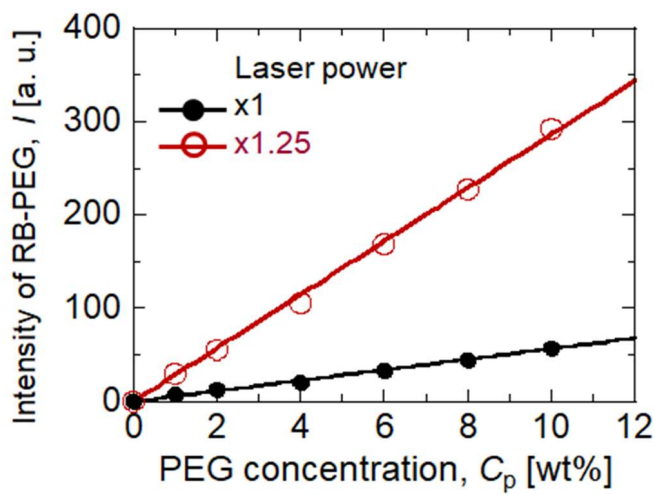

**Figure S1.** Estimation of the concentration of PEG $C_p$ from the fluorescence intensity of RB-PEG. The average fluorescence intensity per unit volume was measured using confocal laser fluorescence microscopy while repeatedly diluting 10 wt% PEG with 0.1 wt% RB-PEG. The data points are the average value of 5 different points, and the solid lines are the linear fit. The color indicates the difference in laser power, the red is 1.25 times higher than the black.

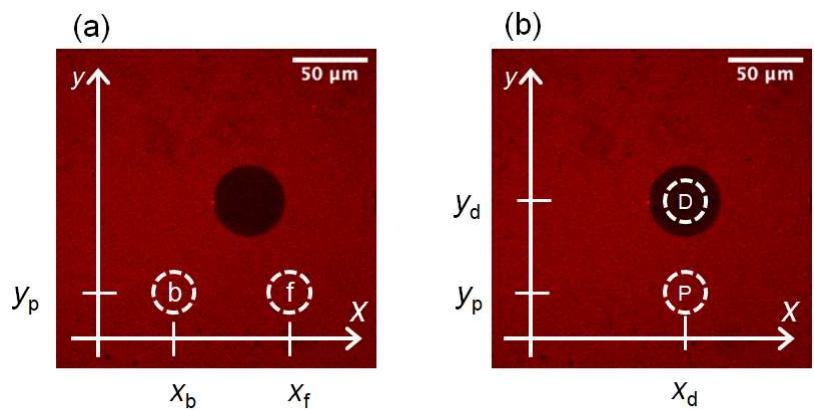

**Figure S2.** Schematic illustration for estimating (a) the PEG concentration gradient in front of and behind the motile dextran-rich droplet and (b) the PEG concentration difference inside and outside the droplet. (a) Average intensities of RB-PEG at the front ($x_f$, $y_p$) and behind ($x_b$, $y_p$) of the motile droplets were used to estimate the PEG concentration gradient along the motile direction, $dC_p/dx$. (b) For estimating the concentration difference of PEG $\Delta C_p$, the average intensity of RB-PEG was measured inside the droplet ($x_d$, $y_d$) and the outside ($x_d$, $y_p$).

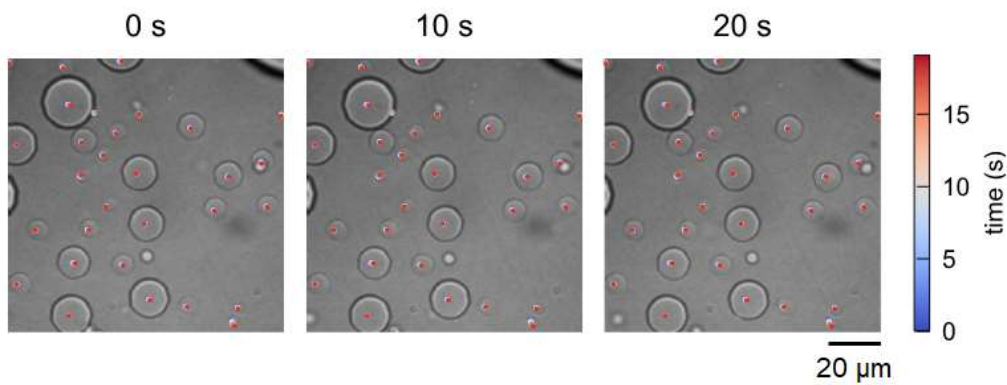

**Figure S3.** Time series of dextran droplets for 20 seconds before the addition of water. The trajectory of the droplet center and its time are shown as colored dots on the right.

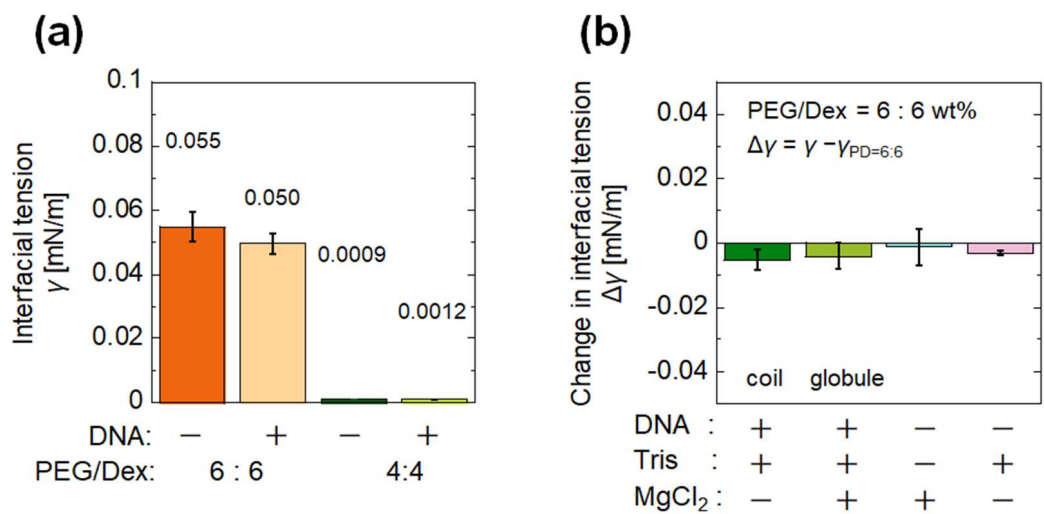

**Figure S4.** (a) Interfacial tension between PEG-rich and dextran-rich phases for PEG/dextran blends (initial compositions of the PEG/dextran (wt%/wt%) are 6:6 and 4:4, respectively) with and without coil DNA. (b) Change in interfacial tension from PEG/ dextran system (initial composition is PEG/dextran (wt%/wt%) is 6:6) with and without DNA, 0.3 mM Tris (pH 8.0), and 20 mM $MgCl_2$. Error bar is standard error (n = 2−5).

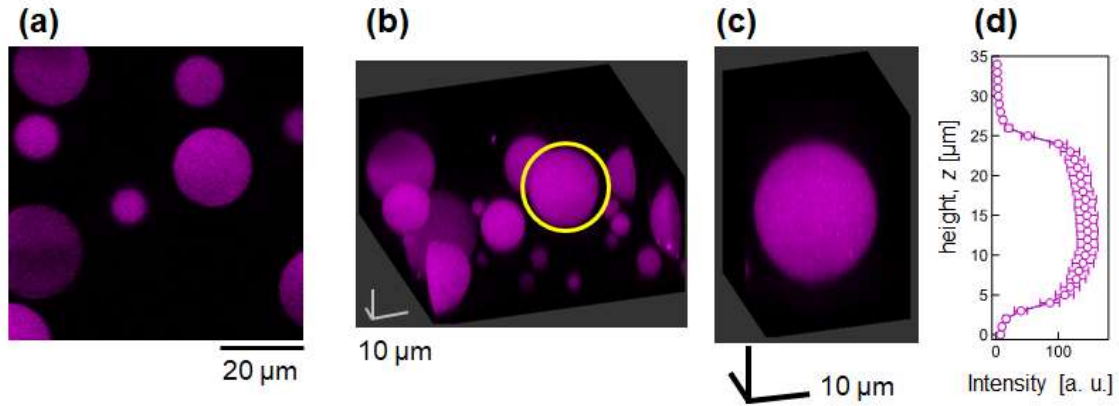

**Figure S5.** Confocal fluorescence microscopy image of dextran droplets containing DNA (shown in purple). (a) Cross-sectional image at the center of the droplet. (b, c) 3D image and magnified view of the droplet surrounded by the solid yellow circle. (d) Fluorescence intensity profile along the z-axis in (c). Error bar is standard error.

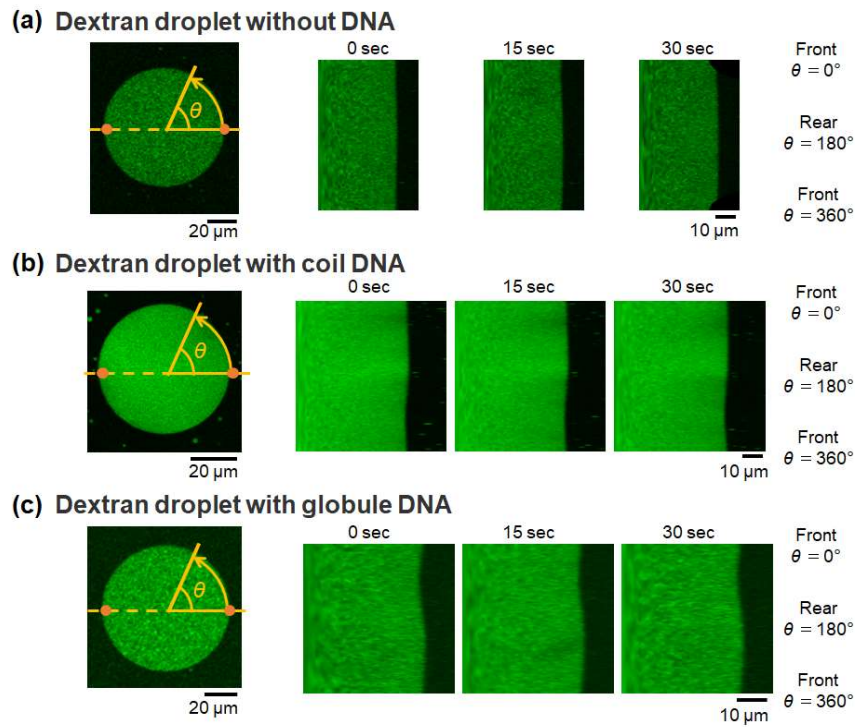

**Figure S6.** Shape of motile dextran droplets with or without DNA. (left) Cross-sectional images of dextran droplets (shown in green). (right) Reconstituted images from the radial profile and the time development, where the front and rear parts of the droplets are shown as $\theta = 0°$, 360° and 180°, respectively. The rear part of the droplets containing DNA (b, c) appears to be slightly convex in shape compared to the other regions.

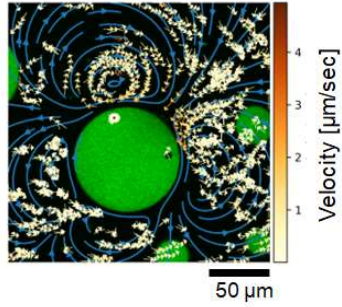

**Figure S7.** Convection inside and outside the motile dextran droplet (shown in green) with coil DNA. The 120 s trajectories of small domains inside and outside of the motile droplet adhered on a glass are shown. The color indicates that the magnitude of the velocity. Blue arrows are fitting lines using the squirmer model of the Brinkman medium ($\beta$ = 2.5, puller).

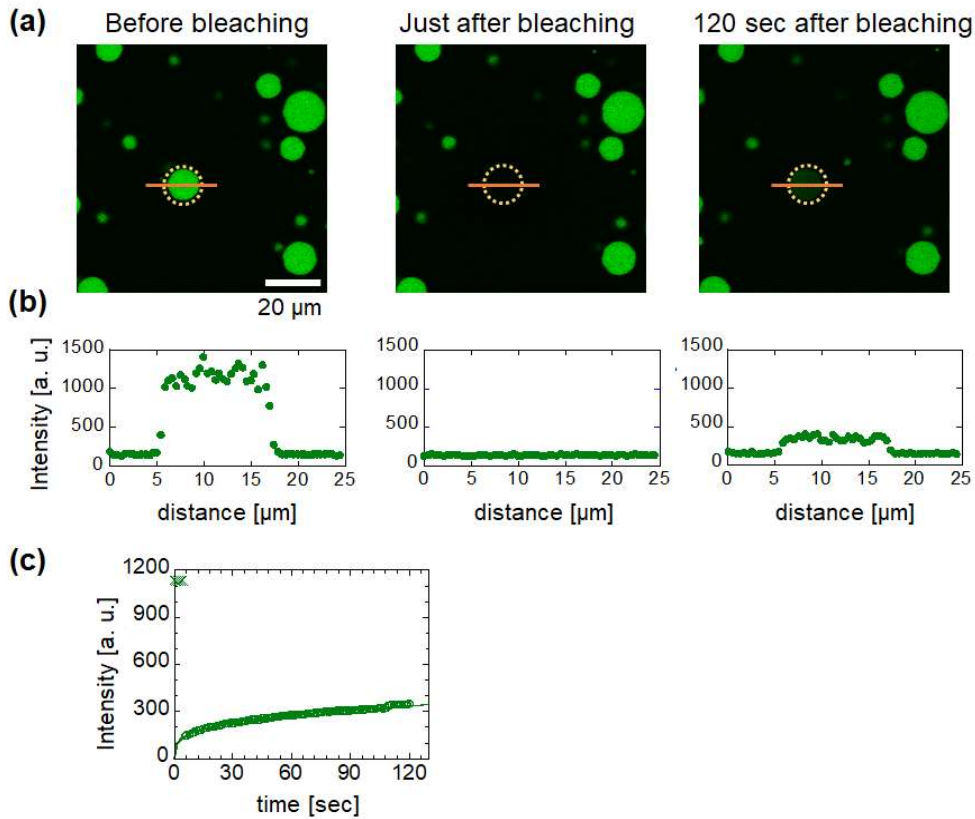

**Figure S8**. FRAP experiment for dextran droplets (shown in green). (a) Time series before and after photobleaching of the target droplet marked by the dotted circle, and (b) fluorescence intensity profile along the solid line on the target droplet in (b). (c) Time course of the average fluorescence intensity of the photobleached region marked by the dotted circle.

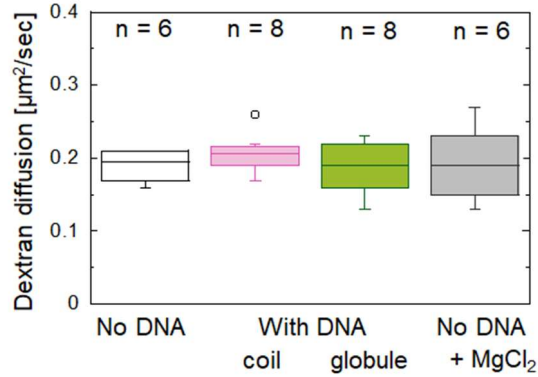

**Figure S9.** Diffusion coefficient of dextran inside dextran droplets measured by FRAP under the no PEG gradient. (From left to right) PEG and dextran solution only, with coil or globule DNA, and with $MgCl_2$, respectively.

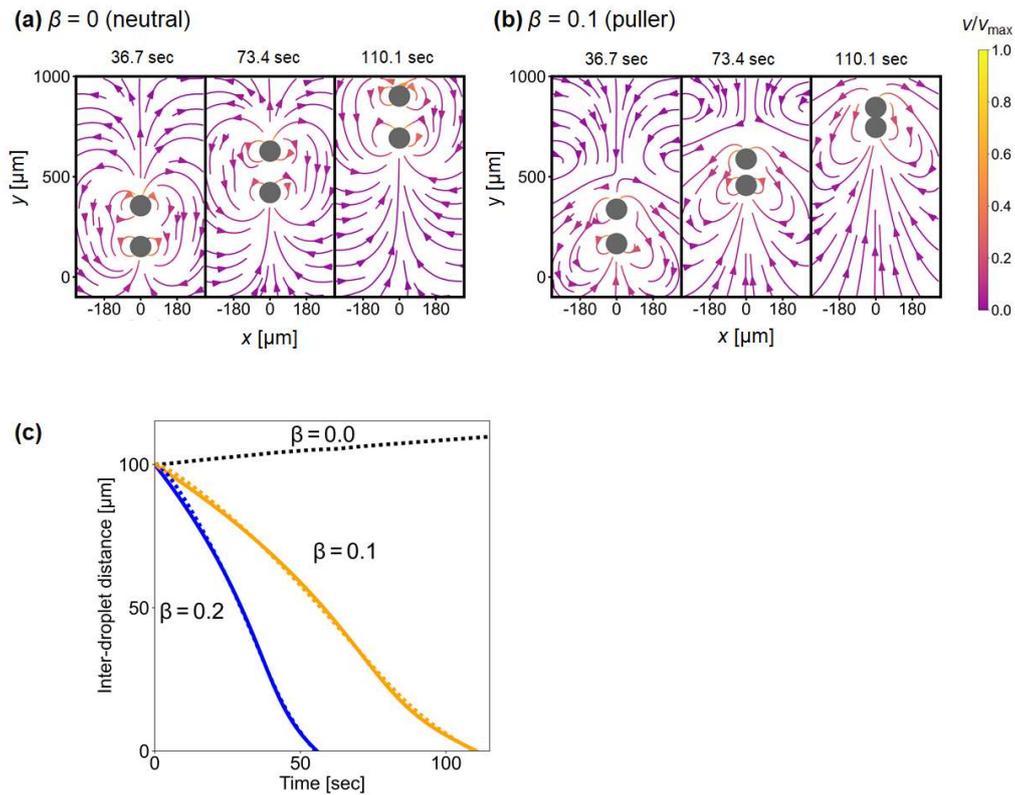

**Figure S10.** (a, b) Snapshots of two motile droplets swimming in the same direction. Swimming types are (a) neutral with $\beta = 0$ and (b) puller with $\beta = 0.1$, respectively. The color of the arrows indicates the magnitude of normalized velocity at the maximum value. (c) Inter-droplet distance (solid line) (dotted line) is plotted for various $\beta$ (dotted line). Solid lines are fitted by with Eqs. (S8, S9).

**Supplemental Movies**

**Movie S1:** Microscopic images of motile dextran droplets moving down a PEG concentration gradient, (from left to right) transmission and fluorescence images of the dextran-rich phase (green) and the PEG-rich phase (red), respectively.

**Movie S2:** Microscopic images of a large dextran droplet stacked on a glass substrate, (from left to right) transmission and fluorescence images of the dextran-rich phase (green) and the PEG-rich phase (red), respectively.